\documentstyle[preprint,aps,prd,epsf]{revtex}
\begin{document} 
\draft 
\tighten

\preprint{\vbox{\hbox{SUSX-TH/97-003}}}

\title{Bubble collisions in SU(2)$\times$U(1) 
gauge theory and the production of non-topological strings.}

\author{ P.M. Saffin\thanks{E-mail: p.m.saffin@sussex.ac.uk}\& 
         E.J. Copeland\thanks{E=mail: E.J.Copeland@sussex.ac.uk}} 
\address{Centre for Theoretical Physics, University of
Sussex, \\Brighton BN1 9QH, United Kingdom}

\date{\today}
\maketitle


\begin{abstract} 
We extend work done previously, \cite{paper}, 
concerning the formation of vortices
in a U(1)$\rightarrow$1 gauge theory at a first order phase transition 
to the symmetry breaking
SU(2)$_{T} \times $U(1)$_{Y}\rightarrow$U(1)$_{Q}$. It is
shown that the collision of bubbles, appearing at the phase
transition, allow the possibility for forming non-topological strings
associated with the gauge group. The method used also shows clearly
how these vortices are related to the Nielsen-Olesen vortices
\cite{no73}.
\end{abstract}
\pacs{98.80.Cq}

\noindent The possible existence of 
vortex solutions within 
the standard model has been realised for some time
\cite{nambu77}-\cite{vach94} and their associated stability has also
been widely studied \cite{vach93}-\cite{goodbandb95}. The very fact that 
these non-topological strings can exist is of interest in its 
own right, but recently a number of authors have speculated about 
their potential role for electroweak
baryogenesis, arguing that 
it is possible to use the local symmetry restoration provided
by the the string to generate a baryon asymmetry 
\cite{davies93}-\cite{nagasawa96}.

Unfortunately arguments about the density of defects formed 
which rely on the 
usual geodesic rule seem to imply that because they are 
non-topological in nature, then very few of them are likely to 
be created at a phase transition 
\cite{nagasawa96}. Essentially, because the vacuum in this
model is a three sphere, a winding around the vacuum of the field at
spatial infinity is not enough to guarantee the vanishing of 
the field at some point, a necessary condition for the existence of a 
string. One also has the disadvantage that a three sphere
vacuum is `larger' than the circular vacuum associated with a U(1) 
symmetry breaking, implying that a winding around the vacuum, which could 
correspond to an electroweak string, will be suppressed.

In this paper we present results on the possible  formation of loops of 
electroweak string in a scenario in 
which the usual geodesic rule breaks down, and argue that 
the implication is the production of electroweak string is 
not as suppressed as has been argued previously. 

In particular we find a class of field configurations that allow the gauge
fields of the non-abelian symmetry group SU(2)$\times$U(1) to be redefined
in such a way that their evolution can be 
described by abelian, U(1), dynamics. Significantly, we find that the process 
of bubble nucleation, thought to be the mechanism 
responsible for the electroweak phase transition 
falls into this class of field structures. This then allows 
us to extend previous results in which we demonstrated how loops 
of topological string could be generated during bubble collisions 
\cite{paper}(see appendix) to the formation of electroweak string loops.

To illustrate the idea we shall first consider the Lagrangian for a gauged 
SU(2) theory,
\begin{eqnarray}
\label{lagrangian} 
{\cal L}&=&\left[(\partial_{\mu}-gW_{\mu})\Phi\right]^{\dagger}
           \left[(\partial^{\mu}-gW^{\mu})\Phi\right] \\ 
\nonumber    &~&-\frac{1}{4}W^{a}_{\mu \nu}W^{a \mu \nu}
              -{\cal V}(|\Phi|),
\end{eqnarray}
\begin{eqnarray}
\label{fields}
W_{\mu\nu}&=&W^{a}_{\mu\nu}t^{a},~~~~~~~~~~
W_{\mu}=W^{a}_{\mu}t^{a},\\
W^{a}_{\mu\nu}&=& \partial_{\mu} W^{a}_{\nu} - \partial_{\nu} W^{a}_{\mu} 
+ g \epsilon^{abc} W^{b}_{\mu} W^{c}_{\nu}, 
\end{eqnarray}
where $\Phi(x)$ is a complex doublet scalar field, and $g$ is the gauge 
coupling constant. This model then has a gauged
SU(2) and global U(1) symmetry, the SU(2) generators are chosen to be 
anti-hermitian and so satisfy,
\begin{eqnarray}
\label{algebra}
\left[t^{a},t^{b}\right]&=&-\epsilon^{abc}t^{c}.
\end{eqnarray}
This is so far very general. However we now look at those fields 
which have a special set of initial configurations, namely those that 
can be written in the form,
\begin{eqnarray}
\label{phiconfiguration}
\Phi(x)&=&\exp\left(\theta(x) n^{a}t^{a} \right)\Psi(x),\\
\label{wconfiguration}
W^{a}_{\mu}(x)&=&0,
\end{eqnarray}
where $n^{a}$ is a constant unit vector, 
\mbox{$\Psi(x)=(0,\rho(x)/\sqrt{2})$} and the time derivatives are also
required to vanish initially. What makes this configuration special 
is that in general we would expect $n^{a}$ to have 
spatial variations, whereas here we have
factored the spatial dependence into the phase 
$\theta(x)$. We may introduce two
new unit vectors, $\underline{\tilde{n}}$ and $\underline{\hat{n}}$ such that
($\underline{n},\underline{\tilde{n}},\underline{\hat{n}}$) form a constant
orthonormal basis. To help understand what our imposed  
initial conditions correspond to,
define the new Lie algebra basis 
$T^{1}=n^{a}t^{a}$, 
$T^{2}=\tilde{n}^{a}t^{a}$,
$T^{3}=\hat{n}^{a}t^{a}$. In (\ref{phiconfiguration}) we 
then see that initially the scalar field 
excitation involves only one generator, $T^{1}$, and 
it may then be expected that
the only gauge field to evolve will be the one associated with this 
generator. In fact as we will see, 
the field equations confirm this expectation and 
it is this fact, that only one gauge 
field evolves, which allows us to extract 
U(1) dynamics from this non-abelian theory.

On substituting (\ref{phiconfiguration}) into the Lagrangian we find 
after some algebra,
\begin{eqnarray}
\label{rhothetalagrangian}
{\cal L}&=&\frac{1}{2}\partial_{\mu}\rho\partial^{\mu}\rho
         +\frac{1}{2}\rho^{2}\partial_{\mu}\frac{\theta}{2}
          \partial^{\mu}\frac{\theta}{2}\\
\nonumber    &~&+\frac{1}{2}\left(\frac{g}{2}\right)^{2}
                 W^{i}_{\mu}W^{i\mu}\rho^{2}
         -\left(\frac{g}{2}\right)\rho^{2} n^{i}W^{i}_{\mu}
         \partial^{\mu}\frac{\theta}{2}\\
\nonumber     &~&-\frac{1}{4}W^{i}_{\mu\nu}W^{i\mu\nu}
         -{\cal V}\left(\rho\right).
\end{eqnarray}
We then define the fields,
\begin{eqnarray}
\label{newgaugefields}
A_{\mu}=n^{i}W^{i}_{\mu},~
H_{\mu}=\tilde{n}^{i}W^{i}_{\mu},~
I_{\mu}=\hat{n}^{i}W^{i}_{\mu},
\end{eqnarray}
and impose the initial conditions on the fields given in 
(\ref{wconfiguration}).

The equations of motion for the vector fields $H_{\mu}$, $I_{\mu}$
involve terms linear, quadratic and cubic in these fields, so one 
solution is that they vanish. However as this solution is
consistent with the initial conditions we impose, it is the one
we are interested in. On substituting the vanishing of these fields
into (\ref{rhothetalagrangian}) one finds the effective Lagrangian,
\begin{eqnarray}
\label{u1lagrangian}
{\cal L}&=&\frac{1}{2}\partial_{\mu}\rho\partial^{\mu}\rho
         +\frac{1}{2}\rho^{2}\partial_{\mu}\frac{\theta}{2}
          \partial^{\mu}\frac{\theta}{2}\\
\nonumber    &~&+\frac{1}{2}\left(\frac{g}{2}\right)^{2}
            A_{\mu}A^{\mu}\rho^{2}
         -\left(\frac{g}{2}\right)\rho^{2} A_{\mu}
          \partial^{\mu}\frac{\theta}{2}\\
\nonumber     &~&-\frac{1}{4}F_{\mu\nu}F^{\mu\nu}
         -{\cal V}\left(\rho\right),
\end{eqnarray}
\begin{eqnarray}
F_{\mu \nu}&=&\partial_{\mu}A_{\nu}-\partial_{\nu}A_{\mu}.
\end{eqnarray}
This can now be immediately 
recognized as a standard Lagrangian with a local U(1) symmetry, 
the U(1) coupling and phase are $g/2$ and $\theta/2$ respectively
(The $\theta/2$ appears due to the $4\pi$ rotation symmetry of SU(2)). In 
other words in the standard notation of a U(1) theory (used in \cite{paper}) 
we have 
\begin{equation}
\label{comparison}
{\cal L}_{\rm SU(2)}(\rho, g, \theta, W_{\mu})
\equiv {\cal L}_{\rm U(1)}(\rho,g/2,\theta/2,n^{i}W^{i}_{\mu}).
\end{equation}
 
It is clear to
see that even though the scalar field transforms under a local SU(2) 
group, there exist Nielsen-Olesen \cite{no73} vortex solutions with 
the effective U(1) gauge field
$A_{\mu}=n^{i}W^{i}_{\mu}$. 
By saying this we mean that the vector field $A_{\mu}(=n^{i}W^{i}_{\mu})$ 
and the modulus, $\rho$, of the Higgs field $\Phi$ 
have the same profile as the Nielsen-Olesen gauge field and Higgs field
respectively, this is not a statement about the topological stability
of these non-abelian vortices. The two parameters required to define 
$\underline{n}$
then define a two parameter family of vortices for the SU(2) Lagrangian.
Thus we have shown 
that if we can find a situation where the initial field configurations 
are of the form given by (\ref{phiconfiguration}), (\ref{wconfiguration}) 
then their dynamics can be described by the equivalent U(1) model, which 
is far easier to determine. On the face of it, it may seem we are demanding 
a great deal, and that such a configuration 
is unlikely to occur. However, we will now demonstrate that these are the 
precise conditions that apply to the nucleation of 
two bubbles of true vacuum in an SU(2) theory.

The initial conditions for a single bubble are described by the 
bounce solution, \cite{coleman77}, where the phase is
constant across the bubble. For the case of two bubbles, as the profile
of the bubble walls falls off exponentially quickly, then if the 
bubbles are suitably separated, the initial conditions
are well approximated by a linear superposition 
of two single bubble solutions.
Each bubble will have its own, constant, phase, hence we write for the two 
bubble configuration (where the bubbles are initially 
separated by a distance $b$)
\begin{eqnarray}
\label{initiala}
\Phi_{initial}(x)&=&\Psi(\underline{x},t=0)\\
\nonumber          &~&+\exp\left(\theta n^{a}t^{a}\right)
                   \Psi(\underline{x}-\underline{b},t=0).
\end{eqnarray}
In (\ref{initiala}) 
we have defined $\Psi(\underline{x},t=0)=(0,\rho_{b}(\underline{x})/\sqrt{2})$ 
where $\rho_{b}(\underline{x})$ is to be the profile of the bounce solution. 
To progress further, we now want to show that this configuration can 
be written in a form similar to that in (\ref{phiconfiguration}). To do 
this write
\begin{eqnarray}
\label{initialb}
\Phi_{initial}(x)&=&\exp\left(\tilde{\theta}(x) N^{a}(x)t^{a}\right)
            {0 \choose \xi/\sqrt{2} },
\end{eqnarray}
where we explicitly allow for the possibility that the unit vector 
$N^{a}(x)$ may have a spatial variation. Equating (\ref{initiala}) and 
(\ref{initialb}) we find that
\begin{eqnarray}
\label{totala}
N^{a}(x)&=&n^{a},\\
\label{totalb}
\xi^{2}(x)&=&\rho^{2}(x)+\rho^{2}(x-b)\\
\nonumber  &~&+2\rho(x)\rho(x-b)\cos(\theta/2),\\
\label{totalc}
\sin^{2}(\tilde{\theta}/2)&=&\left(\rho(x-b)/\xi(x)\right)^{2}
                             \sin^{2}(\theta/2).
\end{eqnarray}
Eqn.({\ref{totala}}) shows that in fact $\underline{N}$ is a constant
vector and thus the $x$ dependence of the phase of the 
whole initial configuration has been
factored out into the $\tilde{\theta}(x)$ so we only need to consider the 
effective U(1) Lagrangian with gauge field $A_{\mu}=n^{a}W^{a}_{\mu}$
for subsequent field evolution. This is useful as it means 
that the results presented in \cite{paper} apply to this situation and 
they imply
that for suitable first order potentials 
it is possible to generate a vortex loop along the intersection
of the two bubbles. This vortex corresponds to a winding around the
vacuum in the direction prescribed by the combination of generators
$n^{a}t^{a}$ and the winding is driven by the effective gauge field 
$A_{\mu}$($=n^{a}W^{a}_{\mu}$).

We now come to describe the process for the case when the global U(1)
symmetry of (\ref{lagrangian}) is elevated to a local symmetry by the
inclusion of an abelian gauge field $B_{\mu}$ via,
\begin{eqnarray}
\label{ewlagrangian} 
{\cal L}&=&\left[{\cal D}_{\mu}\Phi\right]^{\dagger}
           \left[{\cal D}^{\mu}\Phi\right]
         -\frac{1}{4}W^{a}_{\mu \nu}W^{a \mu \nu}\\ \nonumber         
        &~& -\frac{1}{4}B_{\mu \nu}B^{\mu \nu}-{\cal V}(|\Phi|),\\
{\cal D}_{\mu}&=&\partial_{\mu}-gW_{\mu}-\frac{i}{2}g'B_{\mu},\\
W_{\mu\nu}&=&W^{a}_{\mu\nu}t^{a}~~~~~~~~~~
W_{\mu}=W^{a}_{\mu}t^{a},\\
B_{\mu\nu}&=&\partial_{\mu}B_{\nu}-\partial_{\nu}B_{\mu},\\
W_{\mu\nu}&=&\partial_{\mu}W_{\nu}-\partial_{\nu}W_{\mu}
             -g[W_{\mu},W_{\nu}], 
\end{eqnarray}
and $g'$ is the coupling constant associated with the extra U(1) 
gauge field.  
We wish to proceed as before, by looking at initial conditions of the form 
(\ref{phiconfiguration}),(\ref{wconfiguration}) 
and $B_{\mu}=0$, with all time 
derivatives vanishing. Then by defining new vector fields we would like
to obtain an effective U(1) Lagrangian. However, with the introduction of a 
local U(1) gauge field we are including a new generator, the identity, 
into the problem. Once this is included we have a special combination
of generators which annihilate the vacuum to leave us with the residual U(1)
symmetry. It is the existence of this special generator that means we cannot
treat the system with the same generality (i.e. 
arbitrary unit vector $n$ ) we could the pure local SU(2) 
case. We can however look at individual cases and 
to do this we shall work in the 
canonical SU(2) real algebra basis, $t^{i}=\frac{i}{2} \sigma^{i}$, 
where $\sigma^{i}$ are 
the Pauli matrices. In this basis the vacuum defined by 
$\Psi$ in (\ref{initiala}) is annihilated by $t^3 + \frac{i}{2} 1$. 

The important point that we wish to stress is that even in this more 
complicated scenario it is still possible to find a basis in which we 
are able to write the theory as an effective U(1) abelian theory 
analogous to (\ref{u1lagrangian}) but with different combinations of the 
original gauge fields providing the local U(1) gauge field and different 
gauge coupling constants.  

The specific cases we will refer to are
\begin{enumerate}
\item $n^{3}=1$ ($n^{1}=n^{2}=0$)
\item $n^{3}=0$ ($n^{1} = \sqrt{1 - (n^{2})^2}$) 
\end{enumerate}
where the $\underline{n}$ is the unit vector appearing in
(\ref{phiconfiguration}).

Each of these cases require a good deal of algebra to work through them 
which we will not reproduce here. However the results are straightforward 
to describe. 

For case 1. 
we find that the gauge field (the analogue of $A_{\mu}$ 
in (\ref{newgaugefields}) and (\ref{u1lagrangian})) which will create 
a winding of the Higgs field when two bubbles 
collide is nothing other than the field 
associated with the $Z$ vector boson in the standard model. This gives
us a possible mechanism for creating a $Z$ string,  
\begin{eqnarray}
\label{Zboson} 
Z_{\mu}&=&\cos\theta_{W}W^{3}_{\mu}-\sin\theta_{W}B_{\mu},\\
\alpha&=&\sqrt{g^2+g'^2},~~g=\alpha \cos\theta_{W},
\end{eqnarray}
where $\alpha/2$ is the gauge coupling constant. Hence  
the effective U(1) Lagrangian has a phase angle of $\theta/2$
and a coupling constant of $\alpha/2 = g/(2\cos\theta_{W})$.
We stress that all the other components of the gauge fields remain 
zero, they do not evolve dynamically. Thus in the notation introduced 
in (\ref{comparison}) we can write
\begin{equation}
\label{zcomparison}
{\cal L}_{{\rm Z loop}}(\Phi, g, \theta, W_{\mu}, B_{\mu}) 
\equiv {\cal L}_{\rm U(1)}(\rho,\frac{g}{2\cos \theta_{W}},
\frac{\theta}{2},Z_{\mu}).
\end{equation}

For case 2. we find that the 
effective U(1) Lagrangian has $A_{\mu} = n^{a}W^{a}_{\mu}$ being
the relevant vector field that evolves to form a loop of zeros in the
Higgs field at the bubble
collision, just as for the pure SU(2) case described in 
(\ref{newgaugefields}) and (\ref{u1lagrangian}). So, this corresponds 
to the production of loops of $W$ string described earlier in 
(\ref{comparison}). 

We should now address some of the difficulties 
inherent with the formation of non-topological strings.
One issue is that although W and Z string solutions exist, they are
unstable and this could affect the evolution of the (assumed constant)
vector $\underline{n}$. That is to say, if we were 
to take $n^{3}=0(n^{3}=1)$ in order
to generate a W(Z) string, how long does it take for 
perturbations in $\underline{n}$
to grow so that the full SU(2)$\times$U(1) dynamics take over and render
the two special cases considered irrelevant? 
To resolve this issue we undertook
a simulation of the full field dynamics for bubble collisions in the 
SU(2)$\times$U(1) gauge model, (for details see \cite{paper}).
We now look at
the effect of adding a perturbation away from the two cases mentioned above.
Fig~1 shows how the modulus of the Higgs field reacts in a typical
bubble collision. In particular we are interested in that region 
on a path, $\Gamma '$, around the Higgs
zero located at $(x,y)=(0,50)$. A configuration that could produce windings
associated with a W string (case 2 above) 
would be $\underline{n}=(0,1,0)$. To see the effect
of perturbing $n^{3}$, fig~2 shows a slice of Higgs configuration space along
$\Gamma '$ when we initially perturb $n^3$ to $n^{3}=0.01$. 
It shows how the component
$d$ of $\Phi=\frac{1}{\sqrt{2}}{ a+ib \choose c+id }$ 
evolves only a small amount away from zero, 
the value it would have in the unperturbed case $n^{3}=0.0$. In particular 
we see from fig~2 that $|d| < .03$ throughout the path, which should be 
compared to $|\Phi| \sim \sqrt{2}$ in the vacuum. As a guide to understanding 
the figure, we note that the loop lies mainly in the horizontal (c-a) plane 
as is to be expected for this type of $W$ string.
 
From case 1, we know that the condition that may lead to a Z string loop is 
$\underline{n}=(0,0,1)$ and fig~3 shows the effect of initially perturbing 
$n^{2}$ to 
$n^{2}=0.01$. Now we see that the $a$ component of the Higgs field evolves
away from zero, as expected, but only by a small amount ($|a| < 0.02$).
Again, as a guide to visualizing the loop configuration in fig~3, we note 
that the loop lies mainly in the vertical (c-d) plane, the lower component 
of the Higgs. 
An interesting feature of this example is that the Higgs field winds twice
around the vacuum. The reason for this is that the effective U(1) coupling
constant for the Z vector field is larger than that of the effective 
coupling for the W field by a factor of $1/\cos(\theta_{W})$, in this example
we took $g=g'=1.0$. This then allows the
vector field to be more efficient at producing winding, a full description 
of the coupling constant's effect on winding can be found in \cite{paper}.
The conclusion
is that the fields around the loop of Higgs zeros in the collision region
are largely unchanged for small perturbations about the two values 
of $\underline{n}$, namely $n^3=1$ and $n^3=0$.

Another important issue is that of how to characterize non-topological
strings. By their nature they have no gauge invariant topological
number associated with them so one cannot use this to determine their
existence. An analogous situation would be the domain wall solution in
the U(1) model. Given a field configuration how could you deduce if it 
corresponded to a domain wall? The existence of two regions with a phase
differing by $\pi$ is not enough as this does not guarantee the field
vanishes between the regions. 
In the abelian model one has a gauge invariant flux which
can be used to look for the presence of strings, however here again
we find a problem. The standard definition of a gauge invariant flux
involves the tensor $\Phi^{\dagger}W^{a}_{\mu \nu}t^a\Phi$, 
which unfortunately vanishes
for W strings so could not be used to probe for their location. However, 
another 
gauge invariant parameter is the energy, and this may be used to show that
the fields are not gauge equivalent to a vacuum. For the two cases discussed 
above, since the effective U(1) fields around the ring of Higgs zeros 
evolve as in a
U(1) model, the energy distributions will necessarily be the same as for 
the corresponding loop of U(1) string.

We have now illustrated a possible mechanism which allows for the creation of 
loops of non-abelian string in the collision region of 
bubbles of true vacuum. 
Unfortunately, this is not the 
same as showing that 
the string loops would actually form in an electroweak phase 
transition. There are a number of problems to be 
encountered. The first is that the transition 
occurs at a non zero temperature, in a plasma. The conduction currents
that exist in the plasma will couple to the gauge fields, 
reducing their efficacy for vortex formation \cite{paper,kibble}. One also
has to be aware that the phase transition is only weakly first order (if 
at all) so that thermal excitations rather than quantum nucleation could 
become an important mechanism for the field to get into the true vacuum. 
The approach we developed in \cite{paper} is only valid for tunneling 
transitions. 

However, we believe that we have provided a plausible mechanism 
which demonstrates that it 
is possible to have non-topological, 
non-abelian strings form in a first order transition
in a gauge model.
It is remarkable that the dynamics of the theory can be accommodated in a 
much simpler abelian U(1) model when we are considering bubble 
nucleation. 

It is worth commenting on the sizes of the typical loops that are formed.
Because they form in the collision regions of the bubbles, they are 
typically the size of the bubbles themselves when they are formed. This 
is of order the correlation length at formation. 

The above formalism could also be applied to the case
of a global theory admitting bubble nucleation. Recently 
there have been simulations
of a global abelian U(1) model which demonstrated 
that vortex formation is allowed
in the collision of bubbles \cite{srivastava}. Based on the results 
presented here we may then infer that
such vortex production would occur under similar circumstances for a
global symmetry of \mbox{SU(2)$\times$U(1)}. 
Our results are significant because this mechanism of 
forming loops is one that violates the geodesic rule (see \cite{paper} 
for the details). We intend to determine the viability of forming 
electroweak strings by trying to model the electroweak transition in 
a future work. For example it would be interesting to learn the 
number density of such loops that would be formed, and their size 
distribution.

\acknowledgments

We are grateful to M. Goodband, R. Gregory, M. Hindmarsh, R. Holman, 
W. Perkins and T. Vachaspati. PS is grateful to PPARC for financial 
support, and EJC is grateful to Dartmouth College for their hospitality 
during the period when this work was being completed. The work was 
supported by a NATO CRG grant number CRG 930904 and partial support was
also obtained from the European Commission under the Human Capital 
and Mobility program, contract no. CHRX-CT94-0423.
\appendix
\section*{}
In this appendix we shall briefly explain the reason for the breaking 
of the geodesic rule when bubbles collide in an abelian gauge theory 
\cite{paper}.
With hindsight one can readily see why string formation in a first
order phase transition may occur in such a model.
In the U(1) gauge theory we find that the true vacuum solution for the vector
field is $A_{\mu} \propto j_{\mu}$, where $j_{\mu}$ are the components 
of the N\"{o}ther current associated with a global U(1) symmetry.
Now consider the path $\Gamma$ of fig~4. Along AB,at relatively late times, 
the field has settled
into the true vacuum so the N\"{o}ther
current is determined by the gauge field. We also know that the gauge field
will be excited inside one bubble where the boundary of the other bubble
is present, due to the different phases of the two bubbles. These excitations
occur over a length scale determined by the mass of the vector particles
in the true vacuum phase. However, as the scalar 
field does not vary significantly in modulus along AB the only way for
the N\"{o}ther current to match the gauge field variation 
is for $\Phi$ to have a large 
phase variation and, depending on the size of the gauge field, the phase can
wrap entirely around the vacuum manifold. To see if a vortex will be found 
inside $\Gamma$ we need to look at the BCA section of the path. At C, the
scalar field has only just been pulled from the false vacuum and has not
had time to wrap around the vacuum manifold. We conclude then that any 
winding formed along AB will not be undone 
along BCA, so it is possible for the
Higgs field along $\Gamma$ to
form a complete winding around the vacuum, leading to a vortex appearing 
inside the path $\Gamma$.
In the non-Abelian case it may then be expected that a similar mechanism
will be at work, allowing the formation of loops which are characterized
by the zeros of the Higgs field. 


\listoffigures

\noindent Figure 1. A plot of a typical profile of the modulus of the
                    the scalar field in a bubble collision. The path
                    $\Gamma '$ forms a loop enclosing the false vacuum 
                    black region located at $(x,y)=(0,48)$.
\newline
\noindent Figure 2. A slice of the Higgs field around the path $\Gamma '$
                    for a perturbed would be W string 
                    ($n^3 =0.01$ initially). The solid loop represents 
                    the full field $\Phi$ (with the $b$ component projected 
                    out), and the three dotted 
                    loops represent the projections 
                    of $\Phi$ onto the $c-d$, $c-a$ and $a-d$ planes.  
\newline                    
\noindent Figure 3. A slice of the Higgs field around the path $\Gamma '$
                    for a perturbed would be Z string ($n^2 =0.01$ initially). 
                    The solid loop represents 
                    the full field $\Phi$ (with the $b$ component projected 
                    out), and the three dotted loops 
                    represent the projections 
                    of $\Phi$ onto the $c-d$, $c-a$ and $a-d$ planes.  
\newline
\noindent Figure 4. Two colliding bubbles showing the path $\Gamma$ 
                    which encloses a vortex. The two dots represent the 
                    flux lines of winding +1 (black) 
                    and -1 (white).

\begin{figure}[!htb]
\centering
\mbox{\epsfxsize=300pt\epsffile{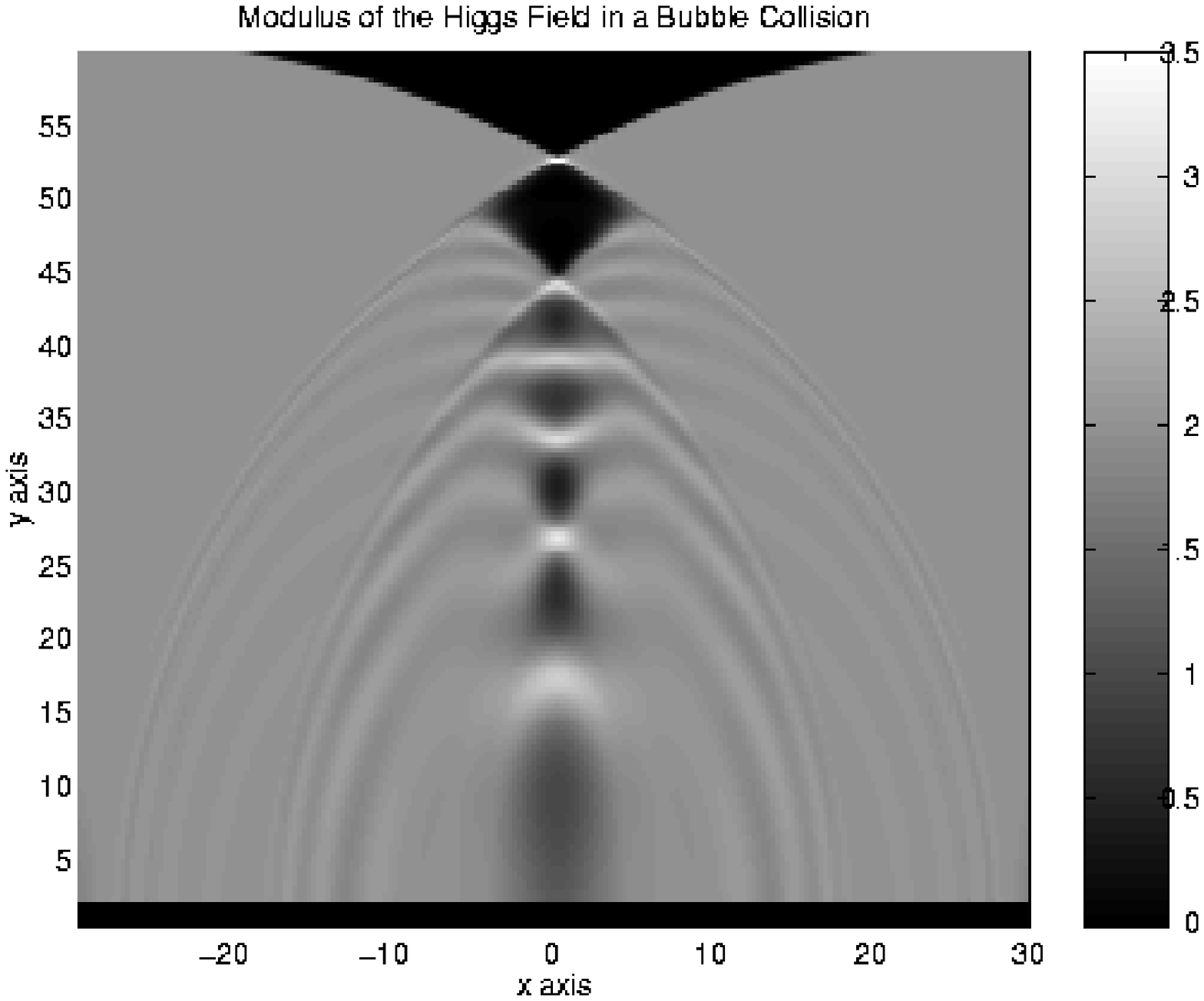}}
	\caption{.}
\end{figure}

\begin{figure}[!htb]
\centering
\mbox{\epsfxsize=300pt\epsffile{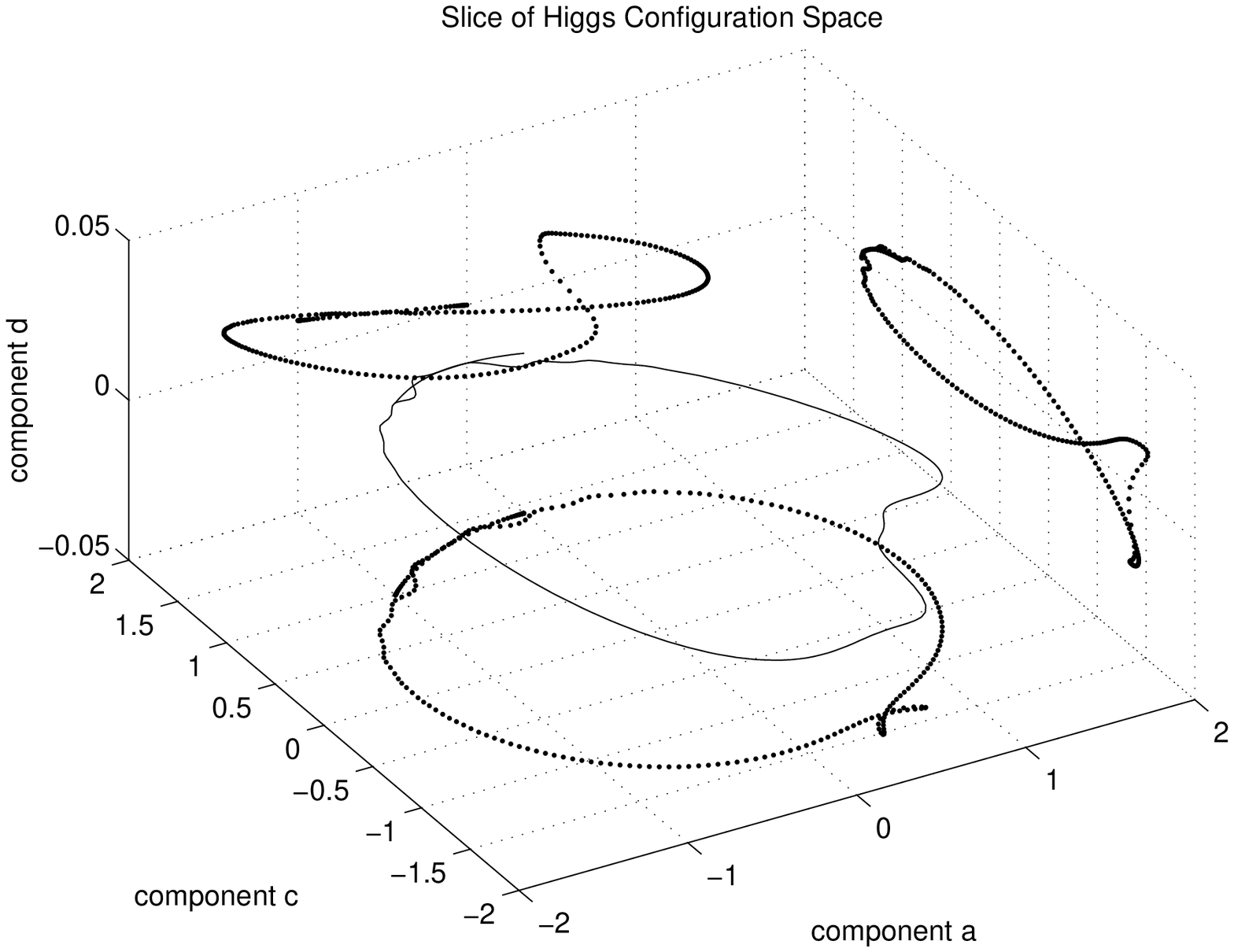}}
	\caption{.}
\end{figure}

\begin{figure}[!htb]
\centering
\mbox{\epsfxsize=300pt\epsffile{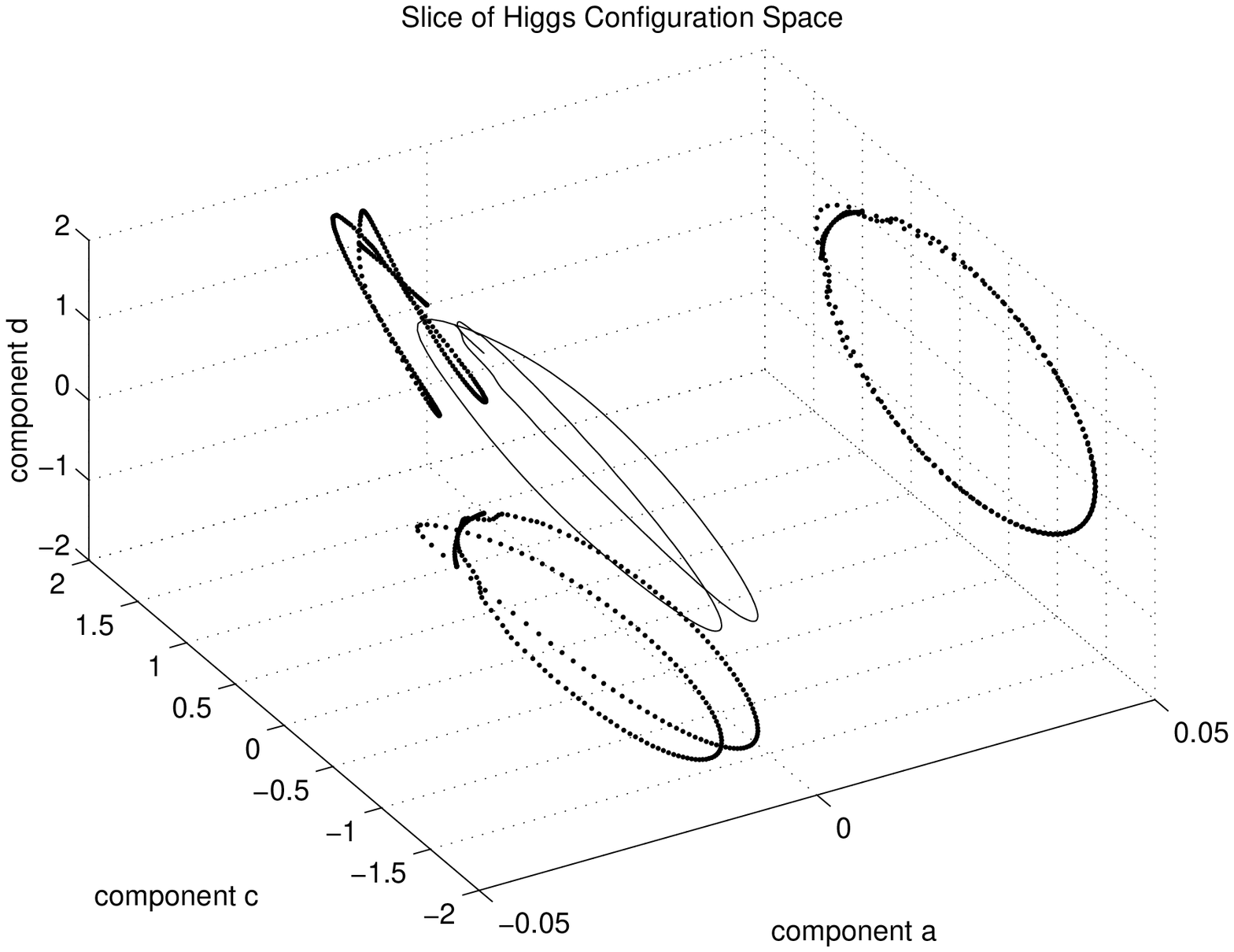}}
	\caption{.}
\end{figure}

\begin{figure}[!htb]
\centering
\mbox{\epsfxsize=200pt\epsffile{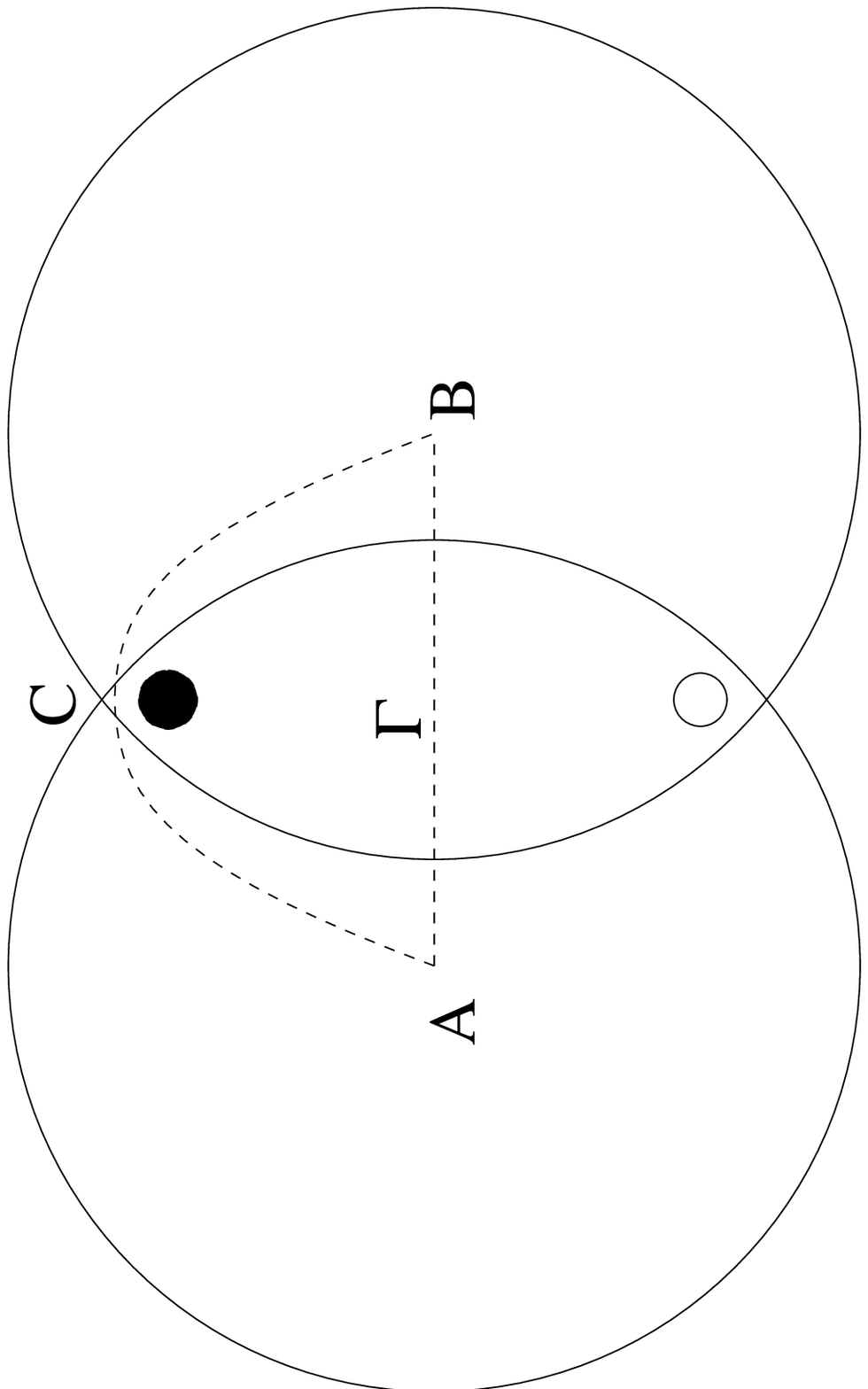}}
	\caption{.}
\end{figure}

\end{document}